\documentclass[oldversion]{aa}
\usepackage{graphicx}
\usepackage{txfonts}
\usepackage{natbib}

      \def\new#1 {{\bf #1 }}
      \def\cut#1 {\sout{#1} }

\def\h      {\ifmmode{^{\rm h}}\else{$^{\rm h}$}\fi}
\def\m      {\ifmmode{^{\rm m}}\else{$^{\rm m}$}\fi}
\def\s      {\ifmmode{^{\rm s}}\else{$^{\rm s}$}\fi}
\def\deg     {\ifmmode{^{\circ}}\else{$^{\circ}$}\fi}
\def\decs  {\ifmmode {\rlap.}$\,$^{s}$\,$\! \else ${\rlap.}$\,$^{s}$\,$\!$\fi}
\def\decas    {\ifmmode{{\rlap.}{''}}\else{${\rlap.}{''}$}\fi}
\def\as     {\ifmmode {\rlap.}$\,$''$\,$\! \else ${\rlap.}$\,$''$\,$\!$\fi}
\def\am     {\ifmmode {\rlap.}$\,$'$\,$\! \else ${\rlap.}$\,$'$\,$\!$\fi}
\def\Msun   {$M_{\odot}$}
\def\kms    {\ifmmode{{\rm km~s}^{-1}}\else{km~s$^{-1}$}\fi}
\def\d {\phantom{$0$}}
\def\dd {\phantom{$00$}}
\def\ddd {\phantom{$000$}}

\begin{document}

\title{The distance to the Orion Nebula}

\author{K. M. Menten\inst{1} \and M. J. Reid\inst{2} \and J. Forbrich\inst{1,2} \and A. Brunthaler\inst{1}}
\institute{Max-Planck-Institut f\"ur Radioastronomie, Auf dem H\"ugel, D-53121 Bonn, Germany\\ \email{kmenten, brunthal@mpifr-bonn.mpg.de}
 \and Harvard-Smithsonian Center for Astrophysics, 60 Garden Street, Cambridge, MA 02138, USA\\  \email{reid, jforbrich@cfa.harvard.edu}}

\date{Received; accepted}

\abstract
{We have used the Very Long Baseline Array to measure the trigonometric parallax of several member stars of the Orion Nebula Cluster showing non-thermal radio emission. We have determined the distance to the cluster to be $414\pm 7$ pc. Our distance determination allows for an improved calibration of luminosities and ages of young stars. We have also measured the proper motions of four cluster stars which, when accurate radial velocities are measured, will put strong constraints on the origin of the cluster.}


\maketitle

\section{Introduction}
The rich Orion Nebula Cluster, the best-studied of all star
clusters, is an important ``laboratory'' for stellar astrophysics and
early stellar evolution.
Most of its $\approx3500$ stars formed within
the past $2~10^6$ yr. The most massive stars, the
$\theta^1$ ``Trapezium" system, have evolved
close to the main sequence \citep{Hillenbrand1997, PallaStahler1999}, while
many lower mass members are still pre-main sequence (PMS) stars.
The dominant Trapezium star $\theta^1$C (spectral type O5--O7) 
is responsible
for the excitation of the spectacular Orion Nebula (ON) Messier 42, the only ionized nebula
visible with the naked eye and extensively studied since the
seventeenth century \citep{Herczeg1998}, since the 1950s at all wavelengths.

The youngest stars in the region are found in the nearby
Becklin-Neugebauer/Kleinmann-Low (BN/KL) region.
The BN/KL region is a very dense part of Orion Molecular Cloud-1,
which is located just a fraction of a parsec behind the ONC \citep{Zuckerman1973, GenzelStutzki1989} and is home to
at least one high-mass star that is currently forming (Orion-I) \citep{Churchwell_etal1987, MentenReid1995, Reid_etal2007}.

Given the importance of the ONC and the BN/KL region for the study of
star formation and early stellar evolution, knowledge of their distance,
$D$, is of great interest.
For example, luminosities are a critical constraint on stellar
evolution models, which deliver stellar masses and cluster ages.
Note that a luminosity determination depends on the square of the distance!
In addition, accurate distances are needed to determine binary-star
separations, to convert angular motions to velocities, and
to estimate stellar mass-loss rates.

Distance estimates for the ONC from the early 1900s ranged from
2000 pc
based on apparent magnitudes and colors of faint stars in
the ON's neighborhood, to 180 pc, based on
the moving cluster method using stars widely distributed around the
ON \citep{Pickering1917, Kapteyn1918}.  Later, luminosity-based distance estimates
of the Trapezium stars were plagued by the poorly known and spatially
variable extinction within the nebula and ranged from 540 pc
to 300 pc \citep{Trumpler1931, Minkowski1946}.
Various distance estimates from the 1960s through the 1980s,
summarized in \citet{Jeffries2007}, used optical and near-infrared (NIR)
photometry and color-magnitude diagrams to yield 347 pc $< D <$ 483 pc.

Uncertain extinction corrections and variations of extinction across
the Nebula, quantified, e.g., by \citet{BohlinSavage1981}, may have led to
part of the large spread in distances published over the last few decades.
Thus, measuring distance by methods that are independent of photometry
is highly desirable.  Comparing radial velocities and proper motions
for samples
of stars for which both could be measured has yielded distances from 380 pc
to 520 pc \citep{Johnson1965,Strand1958}.
Recent modeling of the rotational properties of ONC pre-main-sequence
stars has given $D = 440 \pm 34$ pc for a sample of 74 stars, which
falls to $392 \pm 32$ pc when stars with accretion disks are excluded
\citep{Jeffries2007}. Very Long Baseline Interferometry (VLBI) determinations of the
internal proper motions of ``clouds'' of water vapor (H$_2$O) masers
in the BN/KL region led to another photometry-independent distance
estimate of $D = 480 \pm 80$ pc \citep{Genzel_etal1981}.

Very recently, \citet{Kraus_etal2007} used multiple visual and NIR interferometric observations to determine two orbital solutions for the $\theta^1$C close binary system. One invokes a distance of $434\pm12$ pc and the other $387\pm11$ pc. The existing data do not allow favoring one over the other.

It is apparent from the above summary that there is disagreement
of at least $\pm10$\% among the various methods of estimating the distance to
the ONC, which leads directly to uncertainties of $\pm20$\% for luminosities.
Clearly a much more accurate distance for the ONC would be a fundamental
advance for stellar astrophysics.

The ``gold standard'' for astronomical distance measurements is the trigonometric
parallax, which uses the ancient surveying technique of triangulation.
For astronomical applications, the Earth's orbit is used as the length
scale for one leg of the triangle, and the angular displacement of an
object (relative to very distant sources) directly yields the object's distance.
Optically, only a single ONC member star has a trigonometric parallax,
$361^{+168}_{-87}$ pc obtained from the Hipparcos astrometric satellite
\citep{Bertout_etal1999}; but this measurement is not accurate enough for
most astrophysical applications.
Extraordinarily high astrometric accuracies can now be achieved with
VLBI techniques.  Recently, trigonometric parallax distances to the ON
have been reported to be $437 \pm 19$ pc from water masers in the BN/KL
region using the VERA (VLBI Exploration of Radio Astrometry) array \citep{Hirota_etal2007}
and $389^{+24}_{-21}$ pc from an ONC star (GMR A; see below) using the NRAO\footnote{The National Radio Astronomy
Observatory (NRAO) is operated by Associated Universities, Inc., under a cooperative agreement with the National Science Foundation.}
Very Long Baseline Array (VLBA) \citep{Sandstrom_etal2007}.
These measurements represent a significant advance for the
distance to the Nebula, but the values still differ by $48 \pm 30$~pc.

Using the VLBA, source positions can be now be measured
with an accuracy of $\sim10$ micro-arcseconds ($\mu$as) relative to very
distant quasars.  Recently, VLBA observations of methanol and water masers have
yielded trigonometric parallax distances for the star-forming region W3OH
in the Perseus spiral arm of the Milky Way of 2.0 kpc with $\approx2$\%
accuracy \citep{Xu_etal2006, Hachisuka_etal2006}.
After completing these measurements, we started an astrometric program with the
VLBA in order to determine trigonometric parallaxes for radio-emitting stars
in the ONC.

The Orion Nebula is known to contain about 90 compact radio stars,
either associated with the ONC or with dust-embedded sources in the nearby
BN/KL region \citep{Garay_etal1987, Churchwell_etal1987}.
The emission from many of these sources is thermal in nature,
either resulting from external ionization of circumstellar material
(protoplanetary disks or ''proplyds'') by the UV radiation of
$\theta^1 C$ Ori, or from internal UV sources exciting hypercompact ionized
regions.
However, more than 30 sources have shown significant time variability
and/or emit x-rays \citep{Felli_etal1993, Zapata_etal2004}.
Probably these are non-thermal radio emitters,
making them suitable sources for VLBA parallax measurements\footnote{The VLBA images containing stars presented in this paper 
have an rms noise level around 0.1 mJy~beam$^{-1}$ . In a $2\times1$ milliarcsecond FWHM synthesized beam this corresponds to a brightness temperature of $8.6~10^5$~K.}.


{\tabcolsep=2pt
\begin{table}[t]
\begin{tabular}{lllcccc}
\hline
\hline
Star&
\ddd $\alpha_{\rm J2000}$      &
\ddd $\delta_{\rm J2000}$      &
SIMBAD  & H97     & JW88   & COUP\\
\hline
GMR\\
A  &05 35 11.80318 &$-$5  21 49.2504 & --                  & --      & -- & 450 \\
12 &05 35 15.82615 &$-$5  23 14.1296 &$\theta^1$ A$_1$Ori  & 680     &     &745 \\
G  &05 35 17.95028 &$-$5  22 45.5058 & MT Ori& 815         & 567     & 932 \\
F  &05 35 18.37143 &$-$5  22 37.4342 & V1229 Ori &852     & 589     & 965 \\
\hline
\end{tabular}
\caption{Non-thermally emitting stars in the ONC detected by the VLBA. The first column gives the radio source designation from \citet{Garay_etal1987} and the second and third columns the J2000 position determined by us and calculated for the midpoint of our observations (epoch 2006.46).
The identifications in the fourth through seventh columns come from the SIMBAD database and \citet{Hillenbrand1997, JonesWalker1988, Getman_etal2005}.}

\end{table}
}


\section{Observations and data analysis}
\subsection{\label{obs}Observations}
Our observations with the VLBA were conducted on 2005 September 25,
2006 March 02, 2006 September 09, and 2007 March 6. These dates
well sample the maximum extent of the Earth's orbit as viewed by
the source.  We designed the observations to measure only the
Right Ascension (R.A.) component of the trigonometric parallax
signature, because this component has about twice the amplitude
of the Declination (Decl.) component and the angular resolution of the
VLBA for this source is about two times better in R.A. than in Decl.

The pointing position for the VLBA antennas for the Orion Nebular Cluster was
$(\alpha,\delta)_{\rm J2000} = 05\h35\m15\decs0, -05\deg22'45''$.
We chose the compact extragalactic source J0541$-$054 as the phase-reference
from the International Celestial Reference Frame (ICRF) catalog
\citep{Ma_etal1998}, adopting its position to be
$(\alpha,\delta)_{\rm J2000} = 05\h41\m38\decs083384 \pm 0\decs000019,
 -05\deg41'49\decas42839 \pm 0\as00046$.
We alternately observed J0541$-$054 and the Orion position, switching
sources every 40 seconds over a period of 8 hours during each epoch.
The observing setup used 8 intermediate frequency (IF) bands
of 8 MHz, with 4 detecting right and 4 left circular polarization.
The center frequency of the bands was 8437 MHz and togther they spanned
a total 32 MHz in each polarization.

The data were correlated with the VLBA correlator in Socorro, NM,
with an averaging time of 0.131 seconds.
The position offsets of the ONC stars from the pointing position are small relative
to the FWHM of an individual VLBA antenna's primary beam ($5\am3$).
However, we had to correlate the data in three passes, owing to
the large interferometer fringe-rate differences among them.
Three correlator phase-center positions were used $(\alpha, \delta)_{\rm J2000} = $
$05\h35\m11\decs8022$, $-05\deg21'49\decas229$,
$05\h35\m16\decs2890$, $-05\deg23'16\decas575$, and
$05\h35\m18\decs3706$, $-05\deg22'37\decas436$.
Bandwidth smearing
limits the
effective field-of-view to about
22 
arcseconds about each correlator
position.  Within this field of view,
amplitude decorrelation is less than 10\% on our longest
interferometer baselines.

\subsection{\label{calibration}Calibration}
Most of the data calibration steps are standard and described in
the on-line documentation of the NRAO AIPS software package\footnote{See the AIPS ``cookbook'' under http://www.nrao.edu/aips.}.
Below we discuss additional calibrations that allow improved
astrometric accuracy.

The main source of systematic error for cm-wave phase-referenced
observations is uncompensated interferometric delays introduced by
the Earth's atmosphere and ionosphere, and we conducted supplementary
observations that allowed us to correct for these effects \citep{ReidBrunthaler2004}.
These observations were done by spreading the 8 left-circularly
polarized IF bands to sample about 500 MHz of bandwidth.  Approximately 15 ICRF
sources whose positions are known to better than 1 mas were observed
over a span of about 40 minutes at the beginning, middle and end of
the Orion observations.  The effects of ionospheric delays were removed
by using total electron content values from GPS data.  Also, data were
corrected for the best Earth's orientation parameter values, as the
VLBA correlator by necessity uses preliminary values available at correlation.
Broad-band interferometer residual delays were estimated and modeled as owing to
clock drifts and zenith atmospheric delays.  The phase-referenced data
was then corrected for these effects.

Electronic delay and phase differences among the different IF bands were
removed by measuring them on a strong calibrator, 0530+133, and correcting
the data.   Then the data on the rapid-switching reference source,
J0541$-$054, from all IF bands were combined and tables of antenna-based
interferometer phase versus time were generated.  These phases were
interpolated to the times of the Orion data and subtracted from that data.

All calibrations were applied to the data and synthesis maps were made
using the AIPS task IMAGR.  Typical dirty-beam FWHM sizes were 2.1 by 0.9 mas
elongated in the north-south (NS) direction.  Positions of both the reference source
and the Orion stars were determined by fitting 2-dimensional Gaussian
brightness distributions to the images using JMFIT.

\begin{table*}[t]
\begin{tabular}{lcrrllrrl}
\hline
\hline
Star   &
Epoch &
$S_p$\dd   &
$S$\ddd     &
\ddd $\alpha_{\rm J2000}$      &
\ddd $\delta_{\rm J2000}$      &
$\theta_{\rm maj}$   &
$\theta_{\rm min}$   &
PA \\
& &
(mJy~b$^{-1})$    &
(mJy)&&&
(mas)&
(mas)&
(deg)\\
\hline
GMR A  & 1 & 0.9(0.1)&  0.9(0.3)&05 35 11.803276(6) & $-$05 21 49.24884(20)&$<1.6$&$<0.8$&161(3) \\
       & 2 & 2.1(0.1)&  2.8(0.3)&05 35 11.803031(2) & $-$05 21 49.25059(6) &1.1(3)&0.6(2)&171(30)\\
       & 3 &10.7(0.1)& 13.1(0.2)&05 35 11.803409(1) & $-$05 21 49.25063(2) &1.5(2)&0.5(1)& 170(4)\\
       & 4 &10.8(0.1)& 14.9(0.2)&05 35 11.803135(1) & $-$05 21 49.25219(1) &0.8(1)&0.4(2)&\d 71(8)\\
\\
GMR 12 & 1 & 2.1(0.1)&  6.9(0.5)&05 35 15.826115(5) & $-$05 23 14.12875(9) &3.1(3)& --   &\d58(7)\\
       & 2 &14.4(0.1)& 16.8(0.2)&05 35 15.825945(1) & $-$05 23 14.12943(1) &1.0(1)&0.3(1)&163(3)\\
       & 3 & 4.9(0.1)&  6.8(0.2)&05 35 15.826430(1) & $-$05 23 14.12961(3) &1.6(2)&0.7(1)&173(6)\\
       & 4 & 4.8(0.1)&  5.5(0.2)&05 35 15.826262(1) & $-$05 23 14.13120(2) &&\\
\\
GMR G  & 1 & $<0.24$\ddd	&  \\
       & 2 & 1.1(0.1)& 1.5(0.2) &05 35 17.950071(2) & $-$05 22 45.45702(9) &1.4(4)&$<0.7$&\d19(20)\\
       & 3 & 2.0(0.1)& 2.2(0.2) &05 35 17.950564(2) & $-$05 22 45.45477(7) &$<1.5$& --  &\dd9(40)\\
       & 4 & 4.7(0.1)& 5.1(0.2) &05 35 17.950357(1) & $-$05 22 45.45364(2) &$<0.7$&$<0.6$ &\d34(25)\\
\\
GMR F  & 1 & 1.0(0.1)& 1.9(0.2) &05 35 18.371521(4) & $-$05 22 37.43477(14)&2.1(5)&$<0.9$&162(9)\\
       & 2 & 6.8(0.1)& 7.3(0.2) &05 35 18.371255(1) & $-$05 22 37.43470(2) &$<0.6$&$<0.4$&\d16(43)\\
       & 3 &26.8(0.2)&29.2(0.3) &05 35 18.371659(1) & $-$05 22 37.43379(1) &1.4(1)&$<0.1$&167(3)\\
       & 4 & 2.4(0.1)& 3.1(0.2) &05 35 18.371415(1) & $-$05 22 37.43403(3) &$<1.0$  &$<0.9$&124(35)\\
\hline
\end{tabular}
\caption{Fitted values for the four ONC stars determined by Gaussian fitting. Left to right we list
name, epoch,  peak intensity (brightness), total flux density, R.A., Decl., fitted major and minor axis and position angle. Formal $1\sigma$ uncertainties in the last listed digit returned by JMFIT are given in
parentheses. For R.A. values with smaller uncertainties than 1 $\mu$second, 1 is given for display reasons.  Epochs 1 through 4 correspond to 2005 September 25, 2006 March 2,
2006 September 9, and 2007 March 6.  Absolute position
accuracy is limited the $\pm0.5$ mas ICRF position accuracy of the
phase-reference source, J0541$-$054.  GMR G was not
detected at epoch 1. The quoted brightness is a $3\sigma$ upper limit. For some fits JMFIT did not return a value for the major or minor axis, for some only an upper limit.}
\end{table*}



\subsection{Imaging}
We searched for compact radio emission from a total of 23
stars, but only detected the four stars in Table 1
above a secure ($>5\sigma$) detection threshold of $\approx1$~mJy.
The locations of these stars within the core of the ONC are shown in Fig. 1, using
the stellar designations from Garay, Moran \& Reid \citep[GMR, ][]{Garay_etal1987}.
The star GMR A has no optical counterpart, but has
long been known to be a variable radio source,
whose emission was found by VLBA observations to be
very compact: $\sim1$~milli-arcsecond (mas).
In 2003 January a remarkable outburst of several days duration was
observed at millimeter and radio wavelengths, during which the source's
86 GHz flux density rose more than 10-fold \citep{Bower_etal2003}.
Near infrared (NIR) photometry and spectroscopy indicate that GMR A is a deeply
embedded, very young (T Tauri) star with more than 20 mag of visual extinction.
In x-rays, GMR A is characterized as a ``flare source'' and the
x-ray data indicate an absorbing column of $4~10^{22}$ cm$^{-2}$,
which is consistent with the extinction derived from the NIR data
\citep{Feigelson_etal2002}.
\begin{figure*}[h]
   \centering
   \includegraphics*[width=12cm]{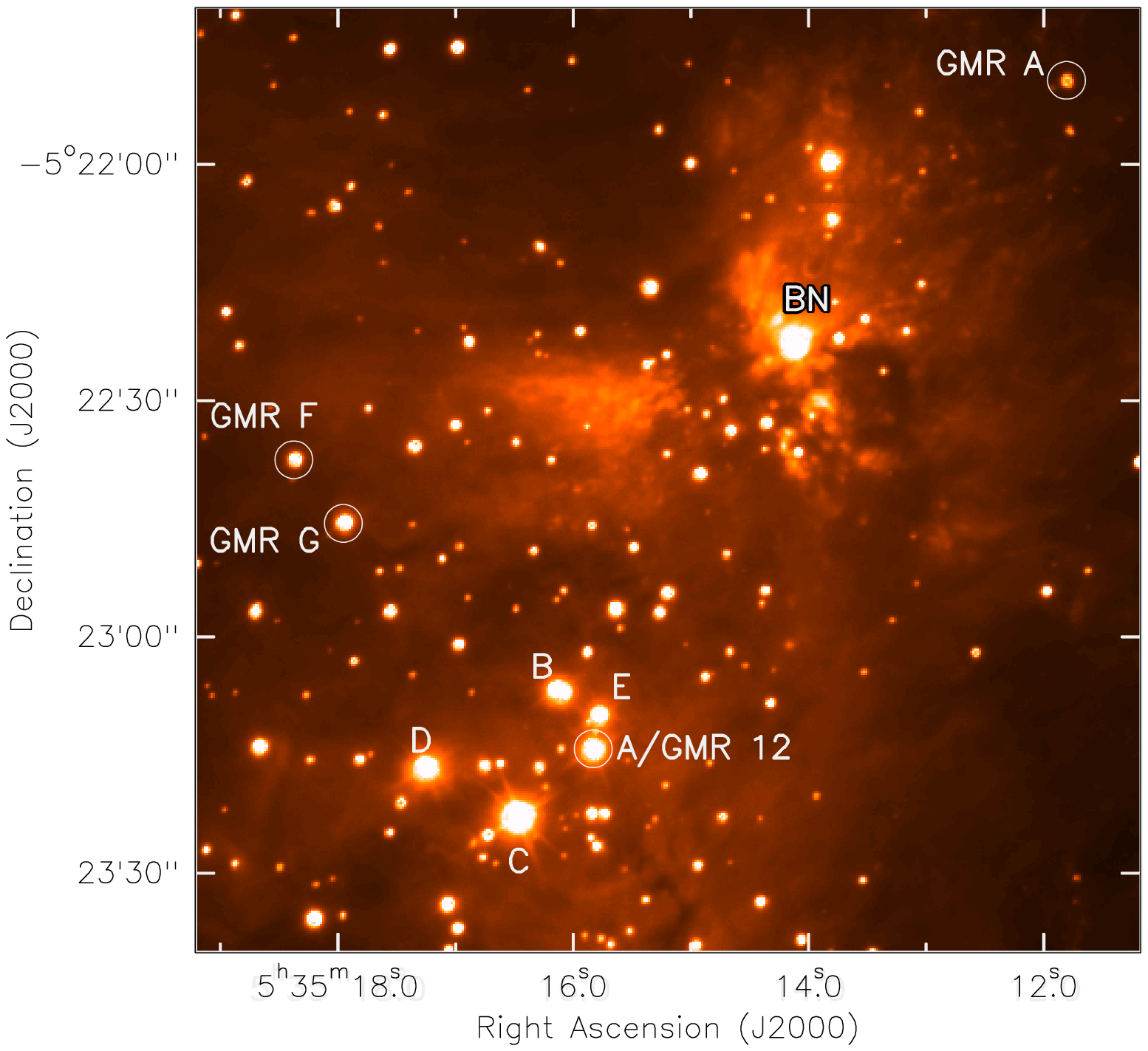}

\begin{center}
\caption{Infrared K-band image of the ONC and the BN/KL region.
(courtesy M. McCaughrean). The Trapezium stars (A, B, C, D \& E) and
the BN object are labeled.  Our program stars (GMR A, GMR F, GMR G \& GMR 12)
are labeled and encircled.  The image was taken with the ESO Very Large Telescope.}
\end{center}
\end{figure*}

The other stars detected, GMR 12, GMR F, and GMR G,
are coincident with optically visible stars and are undoubtedly
ONC members.  The later two stars are listed as $\pi 1925$ and
$\pi 1910$, respectively, in the classic Parenago catalog
\citep{Parenago1954}.  GMR 12 is the binary companion of the Trapezium star
$\theta^1$ A Ori (named $\theta^1$ A$_2$ Ori) .  This $0\as2$ separation binary system
has recently been resolved by NIR speckle interferometry
\citep{Petr_etal1998, Weigelt_etal1999}. (The primary, $\theta^1$ A$_1$ Ori (B0.5 V),
itself is a spectroscopic binary with a cooler companion.)
GMR12 has been detected previously with VLBI observations \citep{Felli_etal1989, Felli_etal1991, Garrington_etal2002}.

Three of our four stars were detected at all four epochs; GMR G
was not detected on the first epoch owing to source variability.
Our observations delivered the positions listed in Table 1 with an
estimated precision of about 0.13 mas in
R.A. and 0.28~mas in Decl. relative to the extragalactic
reference source.  Absolute position uncertainties
are dominated by the $\approx0.5$~mas uncertainty in the ICRF position
of the reference source. Our positions are in excellent agreement with the values determined by \citet{Gomez_etal2005} using the Very Large Array within their larger errors.

Single component JMFIT solutions of our reference source, 0541$-$054,
for each epoch yield an elongated elliptical Gaussian component with
position angles, PAs, between 145 and 158 degrees (east of north) 
that is barely resolved. JMFIT delivers deconvolved sizes between 0.8 and 1.6 mas for the major axis and between 0.3 and 0.4 for the minor axis. This evokes the possibility of a core-jet structure. Note, that to first order, an evolving jet would contribute to the proper motion, but not to the parallax.
Further note that our parallax essentially derives from the east-west (EW) data.

To quantitatively study the effect that source structure might have on the determinations of 0541$-$041's centroid position we conducted 2 (Gaussian) component fits using JMFIT.
In this case the task delivers a very compact stronger component of size $< 0.7$ mas and a weaker elongated component
with a major axis of size $\approx 4$ mas. For the latter's minor axis JMFIT returns upper limits between 0.05 and 0.2 mas
for the first three epochs, while it has broadened to 2 mas in the forth epoch. This ``broadening'' may indicate
more complex source structure emerging.
For the compact component, which almost certainly represents the QSO's core, we found small deviations from the 1 Gaussian fit 
centroid position of $-0.057$, $-0.037$, $-0.064$, and $+0.005$ mas in R.A. direction and $-0.084$, $+0.036$,
$+0.002$, and $+0.002$ mas in Decl. direction for the four epochs, which contribute
to the error floors discussed in \S \ref{results}.
%
%

In Table 2, we present
the stars' positions together with the fitted peak intensities, $S_{\rm p}$,
(brightness values) and integrated flux densities, $S$. A comparison of the numerical
values of  $S_{\rm p}$ and $S$ shows that all sources at most epochs appear only
marginally resolved, if at all. Since the sources appear more resolved when they
have low flux densities and do hardly appear resolved at all when their flux
densities are highest, we ascribe this apparent resolving of sources to residual
calibration uncertainties that affect low signal-to-noise cases relatively the most.

\section{\label{results}Results -- Parallax and proper motion determinations}
We used our positions as input data for a least-squares fitting
program, which modeled the data as the sum of the sinusoidal
parallax term and a proper motion term.  The parallax term
is entirely determined by one parameter ($\pi$), since the
Earth's orbit and source directions are well known.
The proper motion term requires two parameters for each coordinate:
an angular offset and speed.  The first four lines of Table 3
give the best fit parallaxes and proper motions
for all four stars individually. In Fig. 2 we plot the positions of the 4 stars
versus time with the best fit parallax and
proper motion models. Since all stars are at the same
distance, within measurement accuracy, we also present a
combined solution in which we simultaneously solve for a single
parallax parameter and separate proper motions for the three
stars detected in all four epochs (see final line of Table 3 and Fig. 3).

\begin{figure*}
   \centering
   \includegraphics*[width=8.0cm]{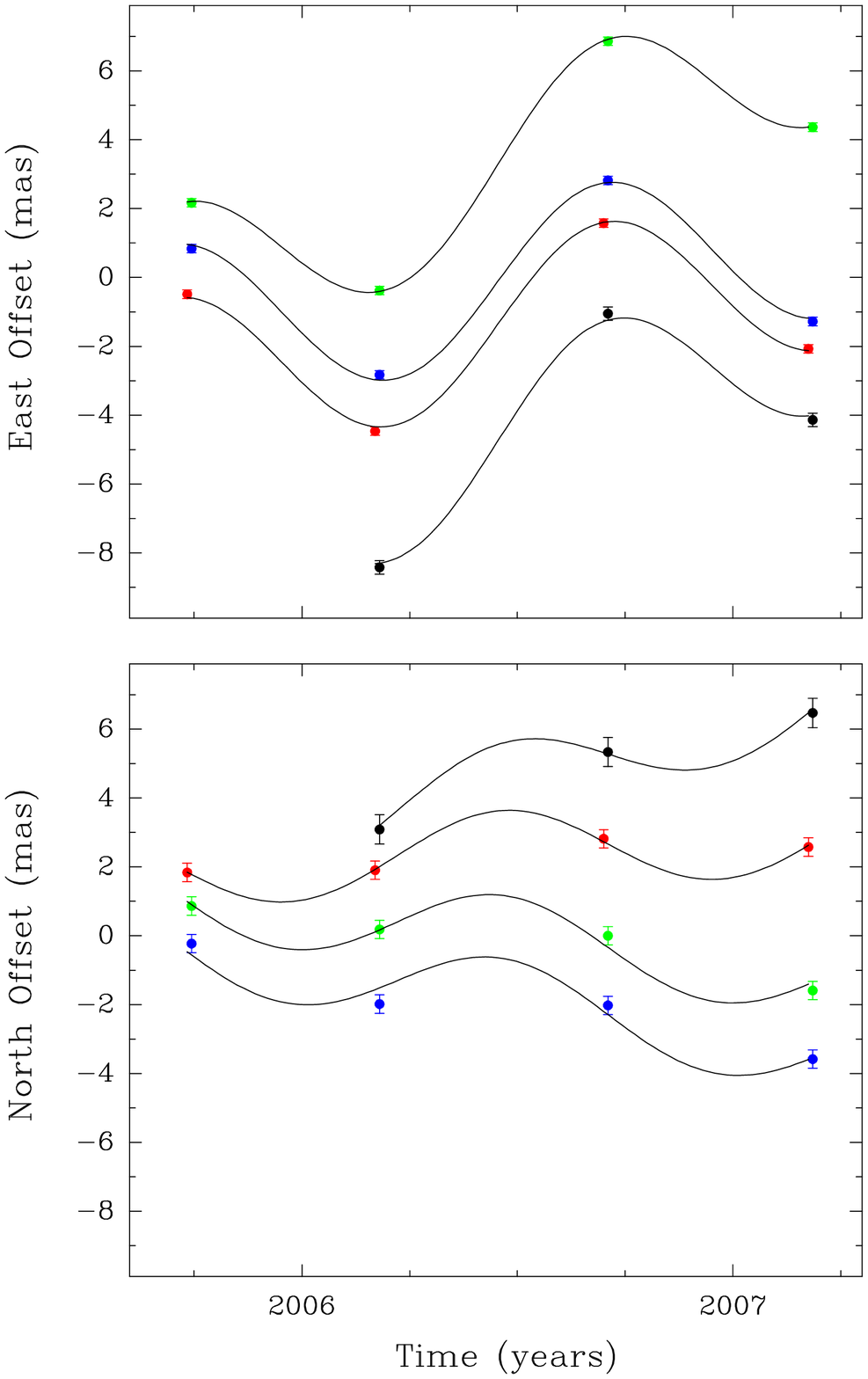}
   \begin{center}
\caption{Position versus time for GMR G (black), GMR F (red), GMR 12(green),
and GMR A (blue).  The top and bottom panels show the eastward
(R.A. cos(Decl.)) and northward (Decl.) offsets, respectively.
For each star the best fit distance from the correlator phase-center
position has been removed and then offset for clarity.
Also plotted are best fit models.  For each star, except GMR G,
the model is characterized by five parameters:
one for the parallax and two for the proper motion speed and
position offset for each coordinate. For GMR G, which was detected at only 3 epochs, the parallax value was fixed at the ``joint-solution'' value
(2.415 mas; see Fig. 3) and only the proper motion parameters were adjusted.
}
\end{center}
\end{figure*}

For this fit, an error ``floor'' was added to the formal position uncertainties in order to account for systematic errors, resulting mostly from unmodeled atmospheric delays and variations in the calibrator positions; see \S \ref{calibration}. The error floors for the R.A. and Decl. data (0.12 and 0.27 mas, respectively) were separately adjusted to bring the reduced $\chi^2$ 
value (per degree of freedom) to unity  in each coordinate. Without the error floors the reduced $\chi^2$ 
values were 3.7 in R.A. and 7.5 in Decl. (per degree of freedom), respectively. After applying these weighting factors, the best fit parallax was $2.415\pm0.040$ mas, corresponding to a distance of $414.0 \pm 6.8$ pc.
Both approaches, the single source average and the joint solution fit, give similar results, and we adopt the joint solution value, i.e., a distance of $414\pm7$~pc.
\begin{figure*}
   \centering
   \includegraphics*[width=8.0cm]{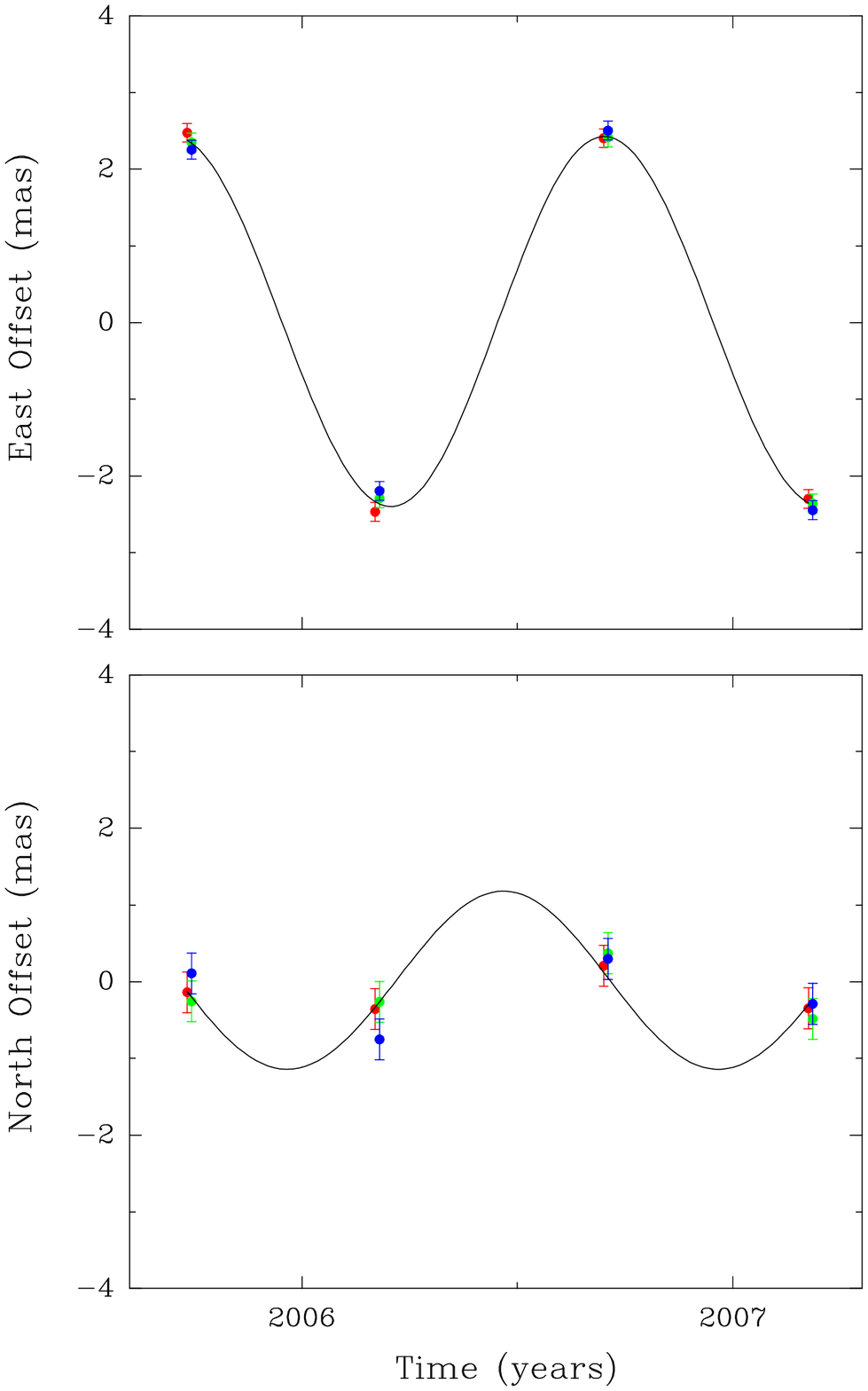}
   \begin{center}
\caption{Joint parallax fit for the 3 stars detected at all 4 epochs using
position versus time data for GMR F (red), GMR 12(green), and GMR A (blue).
The top and bottom panels show the eastward and northward offsets, respectively.
For each star the best fit proper motions have been subtracted.
The best fit parallax was $2.415\pm0.040$ mas,
corresponding to a distance of $414.0 \pm 6.8$ pc.
}
\end{center}
\end{figure*}

We also determined a parallax including the data for GMR G. For this fit, we obtained $\pi = 2.448\pm0.042$ mas, corresponding to $D = 408.5 \pm7.0$ pc. This is within $1\sigma$ with the 414.0 pc determined above, a value which we retain as our best estimate for the reason that of  all sources GMR G consistently showed the lowest flux densities making it the most vulnerable to systematic errors. This is reflected by the fact that the formal error of the fit that included GMR G is not smaller than that of the  three source only fit.

Our distance, 414 pc, is the most accurate measurement for the Orion region, having
an uncertainty of only 1.7\%.
Both of the trigonometric parallaxes mentioned above \citep{Hirota_etal2007, Sandstrom_etal2007} are
consistent with our value within their larger uncertainties of about $\pm20$ pc.

{\tabcolsep=4pt

\section{Discussion}
Our accurate distance is approximately 10\% lower than the 450~pc distance
often assumed for the ONC.   Since luminosity is proportional
to distance squared, our distance implies that
luminosities of the ONC member stars have been over-estimated by nearly 20\%.
Reducing luminosities affects theoretical
investigations of the ONC's star formation history \citep{PallaStahler1999}.
Age and mass estimates from stellar evolution models are discussed
in the literature. We have used an online
tool \citep{Siess_etal2000, Siess2001} which allows one to calculate the change in age and mass, $M$,
for pre-main-sequence stars with masses between 0.1 and 7 \Msun\
and to trace lines for given masses in a luminosity versus temperature
diagram (``PMS tracks'').
For low-mass  ($M <$ 1 \Msun), cool ($<5000$~K) stars, the vertical tracks
indicate little change in mass from a reduced luminosity and we find little
(at most a few percent) difference in mass for a given luminosity.
However, ages are significantly affected and become 20 to 30\%\ larger
by using our accurate distance.
The same happens for higher mass stars, except that masses are
affected, too, and drop by up to 10\%\ for stars with 7 \Msun.

\begin{table}[t]
\begin{tabular}{lllrr}
\hline
\hline
Source&
\dd$\pi$  &
\dd$D$  &
$\mu_{\rm x}$\ddd&
$\mu_{\rm y}$\ddd\\
&
(mas$^{-1}$) &
(pc) &
(mas~y$^{-1}$) &
(mas~y$^{-1}$)\\
\hline
GMR A  &  2.390(0.104) & 418.4(18.2)&  +1.82(0.09) & $-$2.05(0.18) \\ 
GMR 12 &  2.393(0.053) & 417.9(9.2) &  +4.82(0.09) & $-$1.54(0.18) \\ 
GMR G  &   --          &   --       &  +4.29(0.17) & +3.33(0.37)   \\ 
GMR F  &  2.462(0.051) & 406.1(8.4) &  +2.24(0.09) & +0.66(0.18)   \\ 
Mean   &  2.425(0.035) & 412.4(6.0)\\ 
\hline
Joint  &  2.415(0.040) & 414.0(6.8) \\
\hline
\end{tabular}
\caption{Fitted distances and proper motions.
The first column lists the radio source designation.
The second and the third columns give the parallax and distance estimates
derived from a fit to the measured positions in which the parallax and the
proper motions in R.A. and Decl. direction (listed in columns four and five)
were left as free parameters. A proper motion of 1.0 mas~y$^{-1}$ corresponds
to a velocity of 2.0 km~s$^{-1}$ at a distance of 414 pc.
The fifth line gives the weighted mean parallax and distance and their
standard errors of the mean, and the sixth line gives the results of a combined
solution, where only one parallax was fitted to the measurements of
GMR A, GMR 12, and GMR F.}
\end{table}
}

The proper motions of the four stars are in the heliocentric equatorial
reference frame.
We transform them into a Galacto-centric
Cartesian reference frame ($U$, $V$, $W$), where $U$ points toward
the Galactic center, $V$ points toward the direction of Galactic rotation,
and $W$ points toward the North Galactic Pole.  To correct for the Sun's
motion, we used the latest values derived from the analysis of
Hipparcos data \citep{DehnenBinney1998}.
For the distance to the Galactic center
and the circular rotation speed in the Solar Neighborhood,
we assume the IAU values of 8.5 kpc and 220 \kms, respectively.

The calculated $U$, $V$, and $W$
values are dependent on the assumed radial velocities of our stars.
Unfortunately, no accurate radial velocity values have been published for
GMR 12, F, and G.  With the heliocentric radial velocity determined for
GMR A of $+14\pm5$ \kms\ \citep{Bower_etal2003}, we calculate
$U = +9.6 \pm 4.2$ \kms,
$\Delta V = -3.8 \pm 2.6$ \kms\ (slower than Galactic rotation), and
$W = +5.5 \pm  1.7$ \kms,
where $\Delta V$ is the difference between $V$ and the rotation speed
of the Milky Way at the position of the ONC.  For a rotation curve
that is constant with Galacto-centric radius, as appears to be
the case near the Sun, the assumed value of the rotation speed does not
significantly affect $\Delta V$.


If accurate radial velocities for all our program stars could be determined, e.g.,
with IR spectroscopy, such measurements combined with our proper motions could
constrain dynamical scenarios for the formation of the Orion complex,
which is located at the large distance of 19.4 degrees (140 pc) below the
Galactic plane. To explain this puzzling offset one theory invokes a high
velocity cloud colliding with the Galactic plane $6~10^7$ y ago,
sweeping up and compressing material in the process, and subsequently
oscillating about the plane \citep{Franco_etal1988}.

The proper motions of our four stars have a dispersion of 1.5 and
2.6 mas~y$^{-1}$  in R.A. and Decl.
These values, which correspond to 3 and 5 \kms, respectively, are
somewhat uncertain because of small number statistics, but are significantly
larger than velocity dispersions of 1 \kms\ found from optical proper motion
studies \citep{JonesWalker1988, vanAltena_etal1988}.
One explanation for this difference may be
that three or our four stars lie far from the cluster center
and, thus, could be responding to a larger enclosed mass than the
bulk of the stars nearer to the center. 

\acknowledgements{We would like to thank Thomas Preibisch for discussions and Mark McCaughrean for an electronic version of his K-band image. We are grateful to the referee for perceptive comments that helped to improve this paper. Andreas Brunthaler was supported by the Priority Programme 1177 of the Deutsche Foschungsgemeinschaft.}

\bibliography{8247}

\bibliographystyle{aa}







\end{document}